\documentstyle[epsfig,floats,amssymb,aps,prl,twocolumn]{revtex}

\begin{document}

\draft

\title{Coherent Dynamics of Vortex Formation in Trapped Bose-Einstein 
       Condensates}

\author{B.~M.~Caradoc-Davies, R.~J.~Ballagh, and K.~Burnett$^\dagger$}
\address{Department~of~Physics, University~of~Otago, P.~O.~Box~56, Dunedin,
         New~Zealand.\\
         $^\dagger$Clarendon~Laboratory, Department~of~Physics,
         University~of~Oxford, Parks~Road, Oxford~OX1~3PU, United~Kingdom.}

\date{February 5, 1999, revised April 29, 1999}

\wideabs{
\maketitle
\begin{abstract}
Simulations of a rotationally stirred condensate show that a regime of simple
behaviour occurs in which a single vortex cycles in and out of the condensate.
We present a simple quantitative model of this behaviour, which accurately
describes the full vortex dynamics, including a critical angular speed of
stirring for vortex formation. A method for experimentally preparing a
condensate in a central vortex state is suggested.
\end{abstract}
}

The production of vortices has been a central issue in the study of
superfluids. It has been demonstrated for example that attempts to produce bulk
rotation in a cylinder of He~II will lead to vortex production, a state which
calculations show to  be  energetically favoured. The currently realised
\cite{AndersonScienceJul1995,DavisPRLNov1995} Bose-Einstein condensates (BEC)
offer a new medium for studying vortices, and a number of theoretical studies
have considered the properties of static vortices
\cite{DalfovoPRAApr1996,ButtsNatureJan1998}, their stability
\cite{RokhsarPRLSep1997,FetterEprintSep1997,%
FetterEprintAug1998,SvidzinskyPRAOct1998}, excitation spectra
\cite{DoddPRAJul1997,SvidzinskyEprintNov1998},  and phase sensitive detection
techniques \cite{BoldaPRLDec1998}. A variety of methods have  been considered
by which vortices could be formed in a BEC. Fetter \cite{FetterEprintSep1997}
has suggested that a rotating nonaxially symmetric trap could imitate the He~II
rotating cylinder, and obtained an approximate critical rotational speed for
vortex production by a heuristic argument. Jackson {\it
et~al.}~\cite{JacksonPRLMay1998}  have shown that vortices may be generated by
movement of a localised potential through a condensate, while Marzlin and Zhang
\cite{MarzlinPRAJun1998} have calculated vortex production using four laser
beams in a ring configuration. Other numerical simulations, such as collisions
of condensates \cite{BallaghPosterJul1997,JacksonEprintJan1999} have shown in
fact that vortex production appears to be a common consequence of mechanically
disturbing a condensate. 

\begin{figure}
\begin{center}
\epsfbox{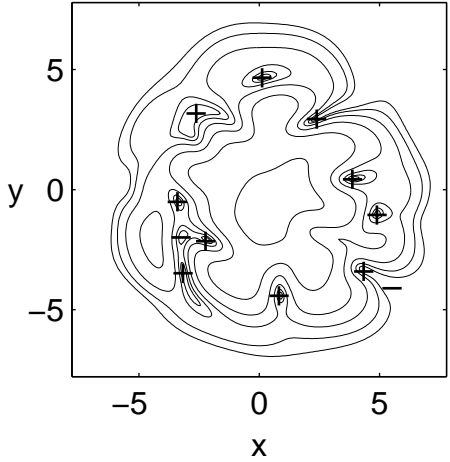}
\end{center}
\caption{Probability density at $t=12\pi$ of a condensate stirred as described
in the text, with the stirrer gradually withdrawn between $t=4\pi$ and
$5\pi$. Contours are logarithmically spaced. Vortices are detected
numerically by searching for their $2\pi$ phase signature, and are marked near
dense regions of the condensate by a $+$ or $-$ sign according to their
sense. $C=88.13$, $\omega_f=1$, $r_s = 3$, $W_0=10$, $w_s=1$.} 
\label{figure:fastend}
\end{figure}

In this paper, we consider a trapped BEC stirred rotationally by an external
potential, and find and analyse a regime where only a single vortex forms. We
present a simple quantitative model of this behaviour, which accurately
describes the full vortex dynamics in terms of a coherent process. The
model gives the critical speed of rotation for vortex formation and explains a
number of other features that are seen, including the stability of a central
vortex (at $T=0$). Our investigation is based on the Gross-Pitaevskii (GP)
equation for the condensate wavefunction $\psi({\bf r},t)$
\begin{eqnarray}
i\frac{\partial \psi ({\bf r},t )}{\partial {t}}&=&-\nabla ^{2}\psi(
{\bf r},t)+V({\bf r},t)\psi ({\bf r},t) \nonumber\\
&&\mbox{}+C|\psi ({\bf r},t
)|^{2}\psi ({\bf r},t),  \label{equation:gpe}
\end{eqnarray}
which is known to accurately describe condensates close to $T=0$. In
Eq.~(\ref{equation:gpe}) we have used scaling and notation as in  Ruprecht {\it
et~al.}~\cite{RuprechtPRAJun1995},  $V({\bf r},t)$ is the external potential,
and $C$ is proportional to the number of atoms in the condensate and the
scattering length. We consider the GP equation in two dimensions only, and
solve it numerically. We simulate the effect of stirring by adding to the
stationary trap potential  a narrow, moving Gaussian potential, representing,
for example, a far-blue-detuned laser \cite{AndrewsScienceJan1997}. $V({\bf
r},t)$ is given  by  $r^2/4+W({\bf r},t)$,  and the stirring potential
\begin{eqnarray}
W({\bf r},t)=W_0\,\mbox{\rm exp}\left[ -4\,\left( \frac{|%
{\bf r}-{\bf r}_{s}(t)|}{w_{s}}\right) ^{2}\right] ,  \label{stirpot}
\end{eqnarray}
is centred at ${\bf r}_{s}(t)$. The initial condensate state for our
simulations is the lowest energy eigenstate of the time-independent GP equation
\cite{DoddPRAJul1997} in which $V$ includes the stationary stirrer.  The
stirrer moves anticlockwise on a circular path, accelerating constantly until
$t=\pi$, when it reaches its final angular speed $\omega_f$.
Figure~\ref{figure:fastend}, which  shows the state of a condensate after it
has been stirred for some time ($t<5\pi$) then left to freely evolve
($t\ge5\pi$),  illustrates the complexity of behaviour that can occur. Vortices
of positive and negative circulation have formed and, as time progresses, move
relative to each other, and annihilate when a positive and negative pair
collide \cite{Website}.

\begin{figure}
\begin{center}
\epsfbox{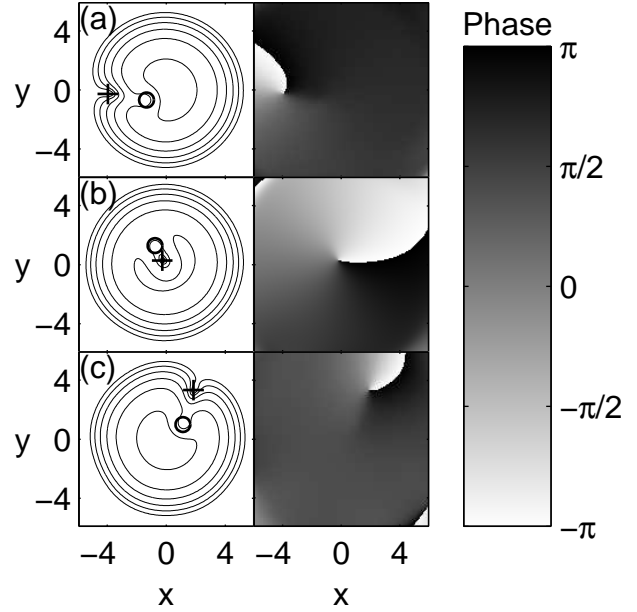}
\end{center}
\caption{Sequence of states for a condensate stirred from rest as described in
the text. Probability density is shown in the left-hand column, and the 
phase
of $\psi$ in the right hand column, for (a) $t=8.80$, (b) $t=18.35$, and
(c) $t=28.15$. The circle denotes the stirrer.
Parameters are as in Fig.~\ref{figure:fastend} except
$\omega _{f}=0.5$ and $r_s=1.5$.}
\label{figure:nearseq}
\end{figure}

\begin{figure}
\begin{center}
\epsfbox{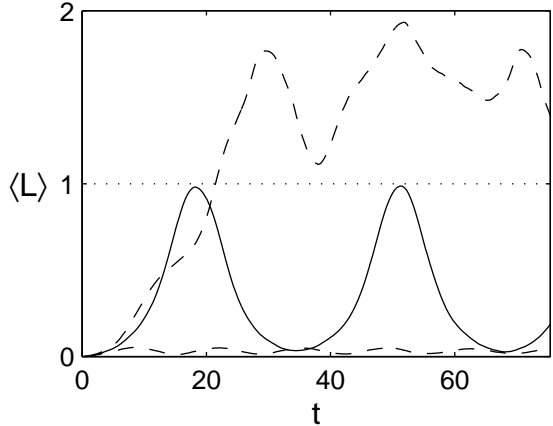}
\end{center}
\caption{Angular momentum expectation values versus time for $\omega _{f}=0.5$
(solid line), $\omega _{f}=0.4$ (lower dashed line), $\omega _{f}=0.6$ (upper
dashed line). Other parameters as in Fig.~\ref{figure:nearseq}.}
\label{figure:angular}
\end{figure}

Amidst the complexity of possible behaviours, an important and simply
characterised behaviour emerges, namely the formation and dynamics of a single
vortex. An example which illustrates the main features is given in
Fig.~\ref{figure:nearseq} where sequential subfigures show the evolution of the
condensate as the stirrer revolves.  A single vortex enters at the edge of the
visible region of the condensate, then cycles to the centre of the
condensate, and back to the edge. This cycle repeats
regularly, as can be seen in Fig.~\ref{figure:angular}  where the solid line
shows the angular momentum ${\langle}L\rangle$ plotted as a function of time
for this case. At lower stirring speeds, similar vortex cycling occurs, but
with progressively smaller amplitudes as $\omega_f$ decreases, so that the
vortex oscillates near the condensate edge. We have found that the condensate
gains angular momentum even for very small values of $\omega_f$. The 
\emph{critical angular speed} which causes a single vortex to cycle right to
the centre of the condensate we denote $\omega_c$. In
Table~\ref{table:omegacrit} we present results from our simulations that show
that $\omega_c$ decreases as $C$ increases, in agreement with the heuristic
result of Fetter \cite{FetterEprintSep1997}. 

The single-vortex behaviour can be understood quantitatively in terms of a
\emph{nonlinear Rabi cycling} model. The essential idea is that the stirring
potential causes the condensate to cycle between the ground state and the first
vortex state, analogous  to the Rabi cycling of an atom in a light field. We 
decompose the condensate on a basis of a ground-state-like part (axially
symmetric) and a vortex part (axially symmetric with an anticlockwise phase
circulation). In the linear (i.e.\ $C=0$) limit the condensate would be
represented  as a superposition of the ground state and the first vortex state
of the trap.  In the nonlinear system, it is more accurate to decompose the
system into collisionally coupled states, in which the radial form of each of
the basis states is modified by its collisional interaction with the other.
Accordingly we assume that the condensate mean-field wavefunction can be
represented approximately as 
\begin{equation}
\psi({\bf r},t) = a_s(t)\phi_s(r,n_v) + a_v(t)\phi_v(r,n_v)e^{i\theta},
\label{equation:ansatz}
\end{equation}
where $r$ and $\theta$ are the cylindrical polar components of ${\bf r}$, and
$n_v=|a_v|^2$. We obtain the lowest energy coupled eigenstates $\phi_s(r,n_v)$
and $\phi_v(r,n_v)e^{i\theta}$ together with their eigenvalues $\mu_s(n_v)$ and
$\mu_v(n_v)$ by solving for a particular value of $n_v$ the coupled
time-independent radial GP equations,
\begin{eqnarray}
\mu_\sigma\phi_\sigma &=& \left[ 
-\frac{1}{r}\frac{d}{dr}\left(r\frac{d}{dr}\right) 
+ \frac{l_\sigma^2}{r^2} \right. \nonumber \\
&& \left. \mbox{}+ \frac{r^2}{4} +
C\left( n_\sigma\phi_\sigma^2 + 2n_\lambda\phi_\lambda^2 \right) \right]
\phi_\sigma.
\label{equation:radialgpe}
\end{eqnarray}
Here $\sigma$ and $\lambda$ are either $s$ or $v$, $l_s=0$ and $l_v=1$ are the
angular momenta of $\phi_s$ and $\phi_ve^{i\theta}$ respectively, $n_v$ is the
fraction of the condensate in the vortex component, and $n_s=1-n_v$ the
fraction in the symmetric component. The $\phi_\sigma$ are  real nonnegative
functions normalized as \mbox{$\int\phi_\sigma^2d{\bf r}=1$}, and $\phi_s$ and
$\phi_ve^{i\theta}$ are of course orthogonal.  The superposition in
Eq.~(\ref{equation:ansatz}) produces a condensate with  angular momentum
expectation value  $\langle L\rangle=n_v$ , and a vortex whose distance from
the centre of the trap decreases as $n_v\rightarrow1$. In the absence of a
stirrer, the vortex  precesses about the centre of the condensate at a
frequency $\mu_v-\mu_s$. Substituting Eq.~(\ref{equation:ansatz}) into
Eq.~(\ref{equation:gpe}),  and projecting alternately onto the states $\phi_s$
and $\phi_ve^{i\theta}$, we obtain a pair of coupled equations for $\dot{a}_s$
and $\dot{a}_v$.
Noting that a constantly rotating stirring potential
$W({\bf r},t)$ can be written \mbox{$e^{-i\omega_ftL}W'({\bf
r})e^{+i\omega_ftL}$}, and writing \mbox{$\tilde{a}_s=a_s e^{i\alpha_s}$}, 
\mbox{$\tilde{a}_v=a_v e^{i(\alpha_s+\omega_f t)}$},  where
\mbox{$\alpha_s(t)=\int_0^t \mu_s(t')dt'$}, we collect the oscillating
exponential time dependences and transform to a frame which rotates with the
stirring potential to obtain the equations 
\begin{mathletters}
\label{equation:rabi}
\begin{eqnarray}
\frac{d\tilde{a}_s}{dt} &=& -i\delta_s(n_v)\tilde{a}_s 
 -\frac{i}{2}\Omega(n_v)\tilde{a}_v, \\
\frac{d\tilde{a}_v}{dt} &=& -i[\Delta(n_v)+\delta_v(n_v)]\tilde{a}_v
 -\frac{i}{2}\Omega^*(n_v)\tilde{a}_s.
\end{eqnarray}
\end{mathletters}
Here $\Delta(n_v) = \mu_v(n_v)-\mu_s(n_v)-\omega_f$ and,
\begin{mathletters}
\begin{eqnarray}
\delta_\sigma(n_v) &=& \int\phi_\sigma(n_v) W'({\bf r})
 \phi_\sigma(n_v) d{\bf r} , \\
\Omega(n_v) &=& 2\int\phi_s(n_v) W'({\bf r}) \phi_v(n_v) e^{i\theta} d{\bf r} .
\end{eqnarray}
\end{mathletters}

Eqs.~(\ref{equation:rabi}) formally resemble the classic Rabi equations, and
hence we identify the $\delta_\sigma$ as frequency shifts and $\Omega$ as the
bare Rabi frequency, but note that here these quantities are variable and
depend on the value of  $n_v$. Despite this nonlinear dependence, the concept
of Rabi cycling provides a simple framework in which to understand the
formation and dynamics of a single vortex: the stirring potential couples and
causes cycling between the initial ground state and the first excited vortex
state. The energy $E'$ in the frame rotating with the stirrer (obtained from
the expectation value of $H'=H-\omega_f L$ where $H$ is the lab frame
Hamiltonian and $L$ is the dimensionless angular momentum operator) is
conserved, and thus in any solution to Eqs. (\ref{equation:rabi}),
$\tilde{a}_s$ and $\tilde{a}_v$ must follow a trajectory that conserves $E'$.
Complete cycling of the vortex to the centre of the condensate occurs when
$n_v$ reaches the value of 1, but this requires the energies in the rotating
frame of the ground state $\phi _{s}(r,n_v=0)$ and the first excited vortex
state $\phi _{v}(r,n_v=1)e^{i\theta}$ to be equal.  Thus the critical angular
speed $\omega _{c}$ is given by the relation 
\begin{equation}
E_{v}-\omega _{c}=E_{g},  \label{equation:omegacrit}
\end{equation}
where $E_g$ and $E_v$ are the lab frame energies of the ground state and first
excited vortex state respectively. A finite stirrer shifts these energies by
$\delta_s(n_v=0)$ and $\delta_v(n_v=1)$ respectively, adjusting $\omega_c$ by
their difference.

In Table~\ref{table:omegacrit} we list the critical angular speeds predicted by
Eq.~(\ref{equation:omegacrit}) for a range of $C$ values, along with the values
of $\omega_c$ found from our numerical simulations of the full GP equation for
$C=30$ and $C=88.13$ cases. The agreement between the predictions from the
two-state model and the full numerical simulation is excellent.

\begin{table}
\begin{tabular}{cdddc}
  $C$    &  $E_g$   &  $E_v$   &  $\omega_c$ & simulation $\omega_c$ \\
\hline
   0     &   1.     &   2.     &  1.         &  -               \\
  30     &   1.811  &   2.520  &  0.709      & 0.6--0.8         \\
  88.13  &   2.744  &   3.284  &  0.540      & 0.5--0.6         \\
 500     &   6.079  &   6.394  &  0.315      & -                \\
5000     &  18.860  &  19.000  &  0.140      & - 
\end{tabular}
\caption{Critical angular frequency $\omega_c$ for the two-dimensional
condensate. The final column gives bounds for $\omega_c$ found from our
simulations of the full GP equation.}
\label{table:omegacrit}
\end{table}

It is difficult to obtain accurate simulations at large values of $C$ for
numerical reasons. The $C=0$ case is easily tractable, and although no visible
vortex cycling occurs below $\omega_f=0.25$, a multiple vortex regime is
entered at $\omega_f=0.8<\omega_c$; the reason being, as Marzlin and Zhang
\cite{MarzlinPRAJun1998} have noted, that the trap levels are equally spaced
for the linear case, so that mixing to higher vortex states readily occurs, and
our two state model is no longer valid. The success of the two state model is
dependent on the fact that for $C\ne0$ the spacing of the levels is
nonuniform. 

\begin{figure}
\begin{center}
\epsfbox{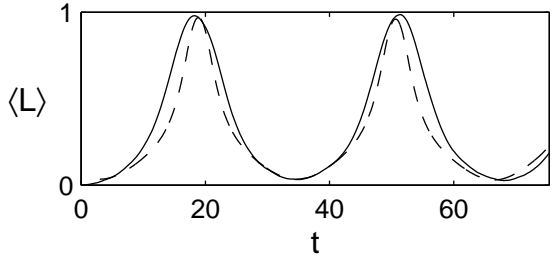}
\end{center}
\caption{Angular momentum versus time for
the full GP equation simulation (solid line) and the two-state model (dashed
line). The two-state model starts at $t=3$ with
\mbox{$\tilde{a}_s=\sqrt{0.965}$} and
\mbox{$\tilde{a}_v=-\sqrt{0.035}$}.
Parameters as in Fig.~\ref{figure:nearseq}.}
\label{figure:twostate}
\end{figure}

The Rabi model also allows us to explain other features of the behaviour, such
as the period of cycling, the response to smaller stirring speeds, and the
effect of different values of stirring radius $r_{s}$ . In 
Fig.~\ref{figure:angular} the condensate response to stirring just below the
critical speed is shown, and reveals an increase in oscillation frequency and
decreased transfer to the pure vortex state, compared to the critical case. In
a simple Rabi model, where the detuning $\Delta $ and Rabi frequency $\Omega $
are constant, the cycling frequency is $\Omega ^{^{\prime }}=\sqrt{\Omega
^{2}+\Delta ^{2}}$, and the maximum value of $n_v$ is $(\Omega /\Omega
^{^{\prime }})^{2}$. By identifying the effective detuning for the two-state
system to be $\Delta+\delta _{v}-\delta_{g}$, these expressions give a
qualitative description of the subcritical stirring in
Fig.~\ref{figure:angular}. A more quantitative treatment requires the nonlinear
character of Eqs.~(\ref{equation:rabi}) to be taken into account, which is
achieved by solving the coupled pair in Eqs.~(\ref{equation:radialgpe}) to find
the eigenvectors and eigenvalues at each value of $n_v$ and then using these to
solve Eqs.~(\ref{equation:rabi}).
We note that in  Eqs.~(\ref{equation:radialgpe}) the term
$2n_\lambda\phi_\lambda^2$ gives rise to an energy barrier between the $n_v=0$
and $n_v=1$ states of the system. The constraint on the system imposed by the
ansatz of Eq.~(\ref{equation:ansatz}) increases this energy barrier slightly
compared to the true (unconstrained) case, and the accuracy of our procedure
can be improved by decreasing this factor of $2$. For example, at $C=88.13$, if
we decrease the factor of $2\rightarrow1.58$, our two-state model produces
behaviour which closely matches the results from the full GP equation, as we
show in Fig.~\ref{figure:twostate}. The energy barrier is deformed by the
presence of the stirrer, allowing the system to cycle between the vortex and
ground state. If the stirrer is far from the centre of the condensate, or is
weak, then $\Omega$ may be too small to distort the energy barrier
sufficiently, and only incomplete cycling occurs even when $\omega_f=\omega_c$.
This feature of the nonlinear system, which agrees with our $r_s=3$ simulations
of the full GP equation, is in contrast to the ordinary Rabi case, where
complete cycling occurs on resonance for any nonzero coupling field.

The validity of the two-state model breaks down when $\omega_f$ exceeds
$\omega_c$, because then higher energy vortex eigen states are energetically
permitted and mixed into the state of the system, as seen, for example, in
Fig.~\ref{figure:fastend} and the upper dashed curve of
Fig.~\ref{figure:nearseq}.

The Rabi model provides some insight into the issue of the stability of a
central vortex state \cite{RokhsarPRLSep1997,SvidzinskyEprintNov1998}. We have
tested this stability in the $T=0$ limit by simulation of the GP  equation,
taking the first excited $l=1$ vortex state and inserting and withdrawing  a
narrow stirrer at a fixed location in the laboratory frame. We find that
although the condensate then wobbles vigorously, the  vortex undergoes only a
very stable small-amplitude oscillation about the trap centre \cite{Website}.
We can interpret this as Rabi cycling of very large effective detuning
(i.e.~$\omega_f=0$), and consequently very small transfer probability out of
the initial vortex state.

The regular cyclic single-vortex behaviour we have found also suggests an
experimental technique for preparing a condenstate in a central vortex state.
By stirring a condensate for a half-cycle, a vortex will be drawn into a nearly
central position.

In conclusion, we have given a simple, quantitative analysis of the single
vortex regime of a stirred condensate. Our two state model captures the
essential coherent dynamics, and accurately predicts the major features of this
regime, but also provides a qualitative understanding in terms of the concepts
of the well known Rabi model. Our result for the critical angular frequency can
be qualitatively related to that for a rotating  cylinder of He~II. However, in
our case the condensate is inhomogeneous, and the  trapping potential plays a
central role, giving rise to well separated condensate eigenstates, of which
only the lowest two become significantly involved. It is worth remarking that
the speed of sound in the vicinity of the perturber has no relevance to the
generation of vortices, in the scenarios we consider here. The model is also
easily generalizable to an arbitrarily shaped stirring potential, including a
rotating anisotropic potential.
Our numerical calculations have been carried out in two spatial dimensions,
but can be expected to apply to ``pancake'' condensates, where the dynamics in
the axial direction are frozen out by very tight axial confinement.
Qualitative features of our results may have even greater generality, since
the two-state model has no direct dependence on dimensionality, and will apply
in
three dimensions if the system symmetry confines the stirrer to couple the
ground state primarily to a single vortex state.

We thank P.~B.~Blakie, C.~W.~Gardiner, and S.~Morgan for valuable discussions.
This work was supported by The Marsden Fund of New Zealand under contract 
PVT603.

\end{document}